\documentclass[12pt,twoside]{article}
\usepackage{amssymb}
\usepackage{amsmath}
\usepackage{latexsym}
\usepackage{epsfig}

\newtheorem{theorem}{Theorem}[section]

\newtheorem{proposition}[theorem]{Proposition}

\newtheorem{conjecture}[theorem]{Conjecture}

\newcommand{\IR}{\mathbb{R}}
\newcommand{\IZ}{\mathbb{Z}}
\newcommand{\IP}{\mathbb{P}}
\newcommand{\IE}{\mathbb{E}}
\newcommand{\IN}{\mathbb{N}}
\begin{document}
\begin{titlepage}
\begin{center}
\vspace*{2,5cm}
 {\Large\bf  Current Fluctuations for the Totally
Asymmetric Simple\bigskip\\
Exclusion Process}
\bigskip\bigskip\bigskip\\
{\large Michael Pr\"{a}hofer and Herbert Spohn}\bigskip\\
Zentrum Mathematik und Physik Department, TU M\"unchen,\\
D-80290 M\"unchen, Germany\medskip\bigskip\\
emails: {\tt praehofer@ma.tum.de}, {\tt spohn@ma.tum.de}\bigskip\\
\today
\end{center}
\baselineskip=24pt \vspace{3cm}\noindent
{\bf Abstract}: The
time-integrated current of the TASEP has non-Gaussian fluctuations
of order $t^{1/3}$. The recently discovered connection to random
matrices and the Painlev\'e II Riemann-Hilbert problem provides a 
technique through which we obtain
the probability distribution of the current fluctuations, in particular
their dependence on initial conditions, and the stationary 
two-point function. Some open problems are explained.

\vspace{1cm}

\end{titlepage}

\setcounter{section}{0}
\section{Introduction} \label{sec.a}
\setcounter{equation}{0}

The asymmetric simple exclusion process (ASEP) has become a paradigm
for nonreversible stochastic particle systems. In this note we
will consider the particular case of one dimension and right jumps
only (TASEP). The TASEP has state space $\{0,1 \}^{\IZ}$ whose
elements, $\eta$, are particle configurations with $\eta_j=0,1$
for $j\in\IZ$. $\eta_j=1$ means a particle at site $j$ and $\eta_j
=0$ means no particle, resp. a hole, at site $j$. The stochastic
dynamics is governed by the following jump rule: independently
each particle jumps with rate 1, i.e. after an exponential waiting
time with mean $1$, to the right neighboring site,  provided
it is empty. Our rule defines the Markov jump process $\eta_t$, $t
\geq 0$. $\eta_0$ is the starting configuration which may be
random.

For the purpose of the introduction let us assume that $\eta_0$ is
distributed according to the Bernoulli measure $\mu_{1/2}$, i.e.
$\eta_{0,j}$, $j\in\IZ$, are independent with $\IP (\eta_{0,j} =
1) = \frac{1}{2}$. Since $\mu_{1/2}$ is a stationary measure for
the TASEP, the process $\eta_{t,j}, t \in \IR, j \in \IZ$, is
stationary in space and time. We are interested in
\begin{equation*}
 N_t = \mbox{ number of particles
which have crossed the bond $(0,1)$ up to time $t$}.
\label{eq:Current}
\end{equation*}
Clearly, $\frac{d}{dt} N_t$ is the current across the origin, hence
our title. By stationarity,
\begin{equation}
\IE(N_t)={\textstyle\frac14}t
\end{equation}
and the real issue are the fluctuations of
\begin{equation}
N_t-{\textstyle\frac14}t\,.
\end{equation}

To convince the reader that the current fluctuations have
something interesting to offer, we first study the large
deviations of $N_t$.
For the lower tail we consider $\IP(N_t \leq at)$ with $a <
\frac{1}{4}$. To reduce the current it suffices to let a single
particle move more slowly. The other particles will then pile up
behind. Therefore we expect
\begin{equation}
\IP(N_t\leq at)\simeq e^{-g_-(a)t}\,,\quad a\leq{\textstyle\frac14}\,,
\end{equation}
for large $t$ with $g_- \geq 0$ and $g_- ({\textstyle\frac14})
=0$. On the other hand for the upper tail, $\IP(N_t \geq at), a >
\frac{1}{4}$, the current must be increased, which requires order
$t$ particles to jump faster. Therefore
\begin{equation}
\IP(N_t\geq at)\simeq e^{-g_+(a)t^2}\,,\quad
a\geq{\textstyle\frac14}\,,
\end{equation}
for large $t$ with $g_+ \geq 0$ and $g_+ ({\textstyle\frac14}) =
0$. Since the large deviations have different order of magnitude
above and below the mean, we must be outside the domain of the
central limit theorem.

A more detailed analysis shows that $g_-(a) \simeq c_{-}|a -
\frac{1}{4}|^{3/2}$ and $g_+ (a) \simeq c_{+}|a -{\textstyle\frac14}|^3$
for $a$ close to $\frac{1}{4}$
\cite{Sep1, Joh}, also \cite{Sep2, Deuzeit} for the PNG model.
Extrapolating beyond the validity of the large deviation result, we conclude 
that for the
lower tail
\begin{equation}
\IP(N_t-{\textstyle\frac14}t\simeq x)\simeq e^{-c_{-}|x/t^{1/3}|^{3/2}}\quad
\mbox{for $x\ll-1$}
\end{equation}
and for the upper tail
\begin{equation}
\IP(N_t-{\textstyle\frac14}t\simeq x)\simeq e^{-c_{+}|x/t^{1/3}|^{3}}\quad
\mbox{for $x\gg1$}\,.
\end{equation}
On this basis the fluctuations should be on the scale $t^{1/3}$ and
\begin{equation}
\label{eq:fluctuations} N_t-{\textstyle\frac14}t\cong t^{1/3}\xi
\end{equation}
for large $t$. $\xi$ is a nondegenerate random variable with a
distribution whose lower tail is $\exp (-c_{-}|x|^{3/2})$ and whose
upper tail is $\exp (-c_{+}|x|^3)$. 

The goal of our note is to explain how (\ref{eq:fluctuations}) and
related quantities like the stationary two-point function,
$\IE(\eta_{0,0} \eta_{t,j}) - \frac{1}{4}$, can be mapped to a last
passage percolation problem with boundary conditions and possibly
defect lines. The asymptotic analysis of such last passage
percolations has been carried out by Baik and Rains \cite{BR} and
we will make contact with their work towards the end.

\setcounter{section}{1}
\section{Last passage percolation with boundary
conditions}\label{sec.b}
 \setcounter{equation}{0}

We generalize the set up of the introduction by taking instead of
$\mu_{1/2}$ the Bernoulli $\mu_{\rho_-, \rho_+}$ as starting
measure, according to which the $\eta_{0,j}$'s are independent
with $\IP (\eta_{0,j} = 1) = \rho_-$ for $j \leq 0$ and
$\IP(\eta_{0,j} = 1)= \rho_+$ for $j \geq 1$. $\mu_{1/2}$ is the
special case with $\rho_- = \frac{1}{2} = \rho_+$. Liggett
\cite{Ligg} determines the law of $\eta_t$ as $t \rightarrow
\infty$. We will come back to his result. Johansson \cite{Joh}
maps the special case $\rho_- = 1, \rho_+ = 0$ to a last passage
percolation. We extend his result to arbitrary $0 \leq \rho_- ,
\rho_+ \leq 1$.

Let us first assign to an ASEP configuration $\eta_{t,j}$ the
height function
\begin{eqnarray}
\label{eq:height}
h_t(j)=\left\{
    \begin{array}{ll}
      2N_t+\sum_{i=1}^j(1-2\eta_{t,i})\,,&{j\geq1}\,,\\
      2N_t\,,&{j=0}\,,\\
      2N_t-\sum_{i=j+1}^0(1-2\eta_{t,i})\,,&{j\leq-1}\,.\\
    \end{array}
\right.
\end{eqnarray}
By fiat $h_0(0) = 0$. $h_t (j)$ is even at even and odd at odd
sites. We will establish that the distribution of the heights can
be obtained from a last passage percolation problem on the
positive quadrant $\IN \times \IN$ with suitable boundary
conditions.

To each site $(i,j) \in \IN_0^2$ we associate a random variable
$w(i,j)$. Let $\zeta_+$ be geometric with parameter $1 - \rho_+$,
$\IP (\zeta_+ = n) = \rho_+ (1- \rho_+)^n$, and independently let
$\zeta_-$ be geometric with parameter $\rho_-$, $\IP(\zeta_- = n)=
(1-\rho_-) \rho_-^n$, $n = 0,1,\ldots$. The $\{w(i,j)|\,(i,j) \in 
\IN_0^2\}$ are independent with distributions:
\begin{eqnarray}
w(i,j)&\mbox{is}&\mbox{\!\!\!exponential with mean $1$,\, $i,j\geq1$}\,,
\nonumber\\
w(0,0)&=&0\,,
\nonumber\\
w(j,0)&=&0\,,\quad 0\leq j\leq\zeta_+\,,
\nonumber\\
w(j,0)&\mbox{is}&\mbox{\!\!\!exponential with mean $(1-\rho_+)^{-1}$,\,
  $j>\zeta_+$}\,,
\nonumber\\
w(0,j)&=&0\,,\quad 0\leq j\leq\zeta_-\,,
\nonumber\\
w(0,j)&\mbox{is}&\mbox{\!\!\!exponential with mean $\rho_-^{-1}$,\,
$j>\zeta_-$}\,.
\end{eqnarray}

Let $\Omega_{m,n}$ be the set of all up/right paths on $\IN_0^2$
starting at $(0,0)$ with end point $(m,n)$. To each path $\omega
\in \Omega_{m,n}$ we associate the passage time
\begin{equation}
T(\omega)=\sum_{(i,j)\in\omega}w(i,j)\,.
\end{equation}
Then the last passage time, from $(0,0)$ to $(m,n)$, is given by
\begin{equation}
G(m,n)=\max_{\omega\in\Omega_{m,n}}T(\omega)\,.
\end{equation}

In the more physical parlance, $\omega$ is a directed polymer with
end points $(0,0)$ and $(m,n)$. Each site carries the energy
$-w(i,j)$. The energy of the polymer $\omega, - T(\omega)$, is the
sum of the site energies along the polymer. Thus $-G(m,n)$ is the
minimal energy, or ground state energy, of the directed polymer.

By construction $G(m,n)$ is nondecreasing in both arguments.
Therefore the level sets of $G$ define a height function which we
denote by $\tilde{h}_t(j)$. More precisely let $A_t =
\{(m,n)|\,\,\,m,n\geq1$, $G(m,n)\leq t \}$ as a random set. $A_t$
is bordered by $\tilde{h}_t$ according to $A_t = \{(m,n) |\,2 \leq m+n
\leq \tilde{h}_t(m-n)\}$.

\begin{theorem}
\label{thm:one}
In the sense of joint distributions we have
\begin{equation}h_t(j)=\tilde h_t(j)\quad\mbox{for}\quad|j|\leq 
h_t(j)\,.
\end{equation}
\end{theorem}
{\it Proof}: The idea is to rotate the TASEP height configuration
by $-\pi/ 4$ relative to the origin and to identify the resulting height 
differences to the right 
as a zero range process. Similarly we rotate the TASEP height
configuration by $\pi/ 4$ relative to the origin and identify the 
resulting
height differences to
the left  as a zero range process. We start with the former.

We define the right dynamics by setting $\eta_{t,0}= 1$ for all $t
\geq 0$. The $\eta_{t,j}$, $j \geq 1$, follow the TASEP rule. For
given $\eta_{t,j}$ let $\zeta_{t,j}$, $j \geq 1$, be the interparticle
distances to the right. Thus if $\eta_{t,1} = 1$, then $\zeta_{t,1}=
0$, since the origin is always occupied. Since the starting
measure is Bernoulli $\rho_+$ for $j \geq 1$, at time $t=0$, the
$\{\zeta_{0,j} \}_{j \geq 1}$ are independently geometrically
distributed with parameter $1 - \rho_+$.
$\zeta_{t,j}$, $j \geq 1$, is a
zero range process where the only allowed transitions are
$(\zeta_{t,j}, \zeta_{t,j+1}) \rightarrow (\zeta_{t,j}+1, \zeta_{t,j+1} -
1)$ for $j = 1,2,\ldots$ and they occur with rate $1$ provided
$\zeta_{t,j+1} > 0$.
$\zeta_{t,1}$ increases in units of 1 and $w(j,0)$, $j = 1,2,\ldots$,
are the successive waiting times of $\zeta_{t,1}$. At $t=0$,
$\zeta_{0,1}$ is
geometrically distributed as $\zeta_+$. Thus $w(j,0) = 0$ for $1
\leq j \leq \zeta_+$. The zero range process $\zeta_{t,j}$, $j \geq
2$, is
stationary. By Burke's theorem \cite{Kel} the current from $2$ to $1$ is Poisson
with intensity $1 - \rho_+$. Thus $w(j,0)$ are independent
exponentials for $j > \zeta_+$.

For the left boundary we only have to interchange particles and
holes. Holes jump to the left only. The left hole dynamics is
defined by setting $\eta_{t,1} = 0$ for all $t \geq 0$. The holes
$1-\eta_{t,j}$, $j \leq 0$, follow the TASEP rule. $\zeta_{t,j}$,
$j \leq 0$, are the interhole distances. $\zeta_{t,0}$ increases
in units of $1$ and $w(0,j)$, $j = 1,2,...$, are the successive
waiting times of $\zeta_{t,0}$. At $t = 0$, $\zeta_{0,0}$ is
geometrically distributed as $\zeta$ and $t \mapsto \zeta_{t,0}$
is a Poisson process with jump rate $\rho_-$.

To build up the random set $A_t$ we start with $\IN^2_0$ such that
$(0,j), 0 \leq j \leq \zeta_+$, and $(j,0), 0 \leq j \leq \zeta_-$
are occupied. For completeness it is useful to regard also the
sites $(j, -1)$, $(-1,j)$, $j = 0,1,\ldots$, as occupied. All
remaining sites are empty. The set of occupied sites defines
$A_0$. The site $(m,n)$ is filled after the random waiting time
$w(m,n)$ starting from the instant of time when both the left
neighbor at $(m-1, n)$ and the lower neighbor at $(m, n-1)$  are
filled. $A_t$ is the set of occupied sites at time $t$. Its
boundary, $\tilde{h}_t$, is defined in the coordinate system
rotated by $\pi/ 4$, to say the coordinates $(m,n)$ are
transformed to the new coordinates $(j,h)$ through $j= m-n$, $h =
m+n$. The distribution of $\tilde{h}_t$ agrees with $h_t$ inside
the cone $\mathcal{C} = \{(j, h)|\,\,h \geq |j|\}$: by
construction the events at the boundary of $\mathcal{C}$ have the
probability induced by the TASEP dynamics. In the interior of
$\mathcal{C}$ the random variables $w(i,j)$ are the waiting times
for the particle jumps of the TASEP.\medskip \hfill $\Box$
 
Returning to the problem of the introduction, we have to set
$\rho_- = \frac{1}{2} = \rho_+$. We define $t \mapsto \tilde{N}_{t}$
as the inverse function to $n \mapsto G(n,n)$. By Theorem \ref{thm:one}
$\tilde{N}_{t}=N_t$ in distribution. Therefore, if one can control the
statistics of the last passage time $G(n,n)$ for large $n$, one
knows the statistics of the current across the origin for
large $t$. More specifically, since $N_t$ and $G(n,n)$ are linear in
average, their fluctuations have the same asymptotic distribution up
to a linear change in scale.

\setcounter{section}{2}
\section{Defect lines}\label{sec.c}
\setcounter{equation}{0}

There are other cases of interest which can be mapped to a last
passage problem. As one example we choose the (deterministic)
initial configuration as $\eta_{0,j} = \frac{1}{2} \big(1 + (-1)^j
\big)$. As before, we want to determine the statistics of $N_t -
\frac{1}{4}t$ for large $t$. The last passage refers now to the
upper triangle $\Delta_n = \{(i,j)|\,\, 1 \leq i, j \leq n, i+j \geq 
n+1
\}$. 
Let $\Omega_n$ be the set of all up/right
paths starting at the anti-diagonal $\{(i,j)|\, i+j=n+1\}$  and
ending at $(n,n)$. Then
\begin{equation}
G^{\mbox{\scriptsize pl}}(n)=\max_{\omega\in\Omega_n}T(\omega)\,.
\end{equation}
 $G^{\mbox{\scriptsize
pl}}(n)$ is the {\it point to line} last passage time, in contrast
to $G (m,n)$ which is the {\it point to point} last passage time.
As before $n \mapsto G^{\mbox{\scriptsize pl}}(2n)$  is the
inverse function to $t \mapsto N_t$. Odd arguments would correspond to 
the initial condition $\eta_{0,j} = \frac{1}{2} \big(1 - (-1)^j
\big)$.

One can force the alternating initial condition into the scheme of
Section 2. We set $\rho_- = 1$, $\rho_+ =0$, and reflect the
$w(i,j)$ at the anti-diagonal, i.e. $w(i,j) = w(n+1-j, n+1-i)$ for
$i+j \neq n+1$. On the anti-diagonal we set $w(i,j) = 0$. Then in
distribution $2 G^{\mbox{\scriptsize pl}}(n)= G(n,n)$, where
$G(n,n)$ is defined as in (\ref{eq:height}). In this scheme we
have point to point last passage with a passage time distribution
symmetric relative to the anti-diagonal.

A second example is the semi-infinite TASEP with a source at the
origin. At site $1$ we insert a particle with rate $\alpha$, $\alpha
> 0$, provided site $1$ is empty. All other jumps are governed by the
TASEP rule. The initial measure is Bernoulli $\rho$. Then the last
passage is restricted to the lower triangle $\{(m,n)|\,n\leq
m\}$. The right boundary of the last passage percolation is
unchanged.
The $w(i,j)$, $j < i$, are exponential with mean 1 and the $w(j,j)$
are exponential with mean $\alpha$. The diagonal is a defect line. To
make the formal analogy even closer to the previous cases we could
copy the $w(i,j)$ in the lower triangle to the one in the upper
triangle, which does not change the statistics of the last passage
time from $(0,0)$ to $(n,n)$.

A widely studied case is the slow bond problem \cite{Janleb,
Sep3}. As initial measure we take $\rho_- = 1$, $\rho_+ = 0$. We
assume that the jumps through the bond $(0,1)$ occur with rate
$r$. This corresponds again to a defect line along the diagonal,
where $w(j,j)$ is exponential with rate $r$. In contrast to the
semi-infinite system the $w(i,j)$'s for $i < j$ and for $i > j$ are
now independent.
One would like to know the average current for $t\to\infty$,
$j_{\infty}(r)=\lim_{t\to\infty}N_t/t$. Clearly, $j_{\infty}(r)= r$ for small
$r$ and $j_{\infty}(r) = \frac{1}{4}$ for $r \geq 1$. The critical
rate $r_c$ is the smallest with $j_{\infty}(r_c) = \frac{1}{4}$.
One knows that $\frac{1}{2} < r_c \leq 1$ and conjectures $r_c =
1$.
In terms of the directed polymer, at $r_c$ there is a depinning
transition. For small $r$ the directed polymer with end points
(1,1) and $(n,n)$ will stay order 1 close to the diagonal. As $r$
is increased, the size of excursions away from the diagonal
increases. For $r > r_c$ the transverse fluctuations diverge 
as $n^{2/3}$ \cite{joh}. The directed polymer depins from the
diagonal.

\setcounter{section}{3}
\section{The two-point function}\label{sec.d}
\setcounter{equation}{0}

The TASEP with Bernoulli $\mu_{\rho}$ $(\rho_+ = \rho = \rho_-)$ as
starting measure is stationary in space-time. From a statistical
mechanics point of view the central quantity is the two-point
function
\begin{equation}
S(j,t)=\IE_\rho\big(\eta_{t,j}\eta_{0,0}\big)-\rho^2\,.
\end{equation}
We list a few properties:
\begin{proposition}
\label{prop:two} We have $S(j,t) \geq 0$, $\sum_j S(j,t) =
\rho(1-\rho)$, $\sum_j j S(j,t) = \rho (1-\rho) (1-2\rho) t$, and
\begin{equation}
\label{eq:laplace}
8S(j,t)=\Delta\big(\IE_\rho\big(h_t(j)^2\big)-\IE_\rho\big(h_t(j)\big)^2\big)
\end{equation}
with the discrete Laplacian $(\Delta f)_j = f_{j+1}-2f_j +
f_{j-1}$ and $h_t(j)$ defined in (\ref{eq:height}).
\end{proposition}
{\it Proof}: $S(j,t)/ \rho(1-\rho)$ is the transition probability
for a second class particle starting at the origin \cite{Ferr},
which implies the first two assertions. For the third one we note
that
\begin{eqnarray}
\frac{d}{dt}\sum_jjS(j,t)&=&\sum_jj\IE_\rho
\big((\eta_{-t,0}-\rho)
(\eta_{0,j-1}(1-\eta_{0,j})-\eta_{0,j}(1-\eta_{0,j+1}))\big)
\nonumber\\
&=&\sum_j\IE_\rho
\big((\eta_{-t,j}-\rho)\eta_{0,0}(1-\eta_{0,1})\big)
=\sum_j\IE_\rho
\big((\eta_{0,j}-\rho)\eta_{0,0}(1-\eta_{0,1})\big)
\nonumber\\
&=&\rho(1-\rho)(1-2\rho)\,,
\end{eqnarray}
where we used conservation of the number of particles.

To see the fourth identity we compute
\begin{eqnarray}
\Delta\IE_\rho \big(h_t(j)^2 \big)&=&\IE_\rho \big
(h_t(j+1)^2-2h_t(j)^2 +h_t(j-1)^2 \big)
\nonumber\\
&&\hspace{-2cm}=-8\IE_\rho
\big(N_t(\eta_{t,j+1}-\eta_{t,j})\big)+2 -4\IE_\rho \big
((1-2\eta_{0,j})(\eta_{0,j+1}-\eta_{0,j})\big)\,,
\end{eqnarray}
using (\ref{eq:height}) and stationarity. By translation
invariance
\begin{equation}
\label{eq:transinv} \IE_\rho \big
(N_t(\eta_{t,j+1}-\eta_{t,j})\big)=\IE_\rho \big
((N_t^--N_t)\eta_{t,j}\big)\,,
\end{equation}
where $N^-_t$ is the number of particles jumping through the bond
$(-1,0)$ up to time $t$. The conservation law ensures that
\begin{equation}
N_t^-+\eta_{0,0}=N_t+\eta_{t,0}\,.
\end{equation}
Inserting into (\ref{eq:transinv}) yields
\begin{equation}
\IE_\rho \big(N_t(\eta_{t,j+1}-\eta_{t,j})\big)= \IE_\rho \big
((\eta_{t,0}-\eta_{0,0})\eta_{t,j}\big)
\end{equation}
and taking into account that $\IE_{\rho}\big(h_t(j)\big) = 2
\rho(1-\rho) t + (1-2 \rho)j$ we obtain (\ref{eq:laplace}).\medskip
\hfill $\Box$

$(\rho(1-\rho))^{-1} S(j,t)$ is a probability distribution with
mean $(1-2 \rho)t$. One expects \cite{Bei Ku Spo} that its variance grows as
$t^{4/3}$ which suggests the scaling form
\begin{equation}
\label{eq:scalingform}
S(j,t)\cong\rho(1-\rho)\big(4(\rho(1-\rho))^{1/3}t^{2/3}\big)^{-1}
{\textstyle\frac18}\,g''\big((j-(1-2\rho)t)
\big(4(\rho(1-\rho))^{1/3}t^{2/3}\big)^{-1}\big)\,.
\end{equation}
The link to the last passage percolation is provided by
(\ref{eq:laplace}) and Theorem \ref{thm:one}, which tell us
that the
two-point function is given through the second moment of the
last passage time at boundary
conditions  $\rho_- = \rho$, $\rho_+ = \rho$.   Because of the subtracted mean in
(\ref{eq:scalingform}) we  consider the passage from
(0,0) to $(m,n)$ for large $n$ with fixed slope $\theta (\rho) =
n/m = (1 - \rho(1-\rho) - (1-2 \rho))/ (1 - \rho(1-\rho)+(1-2 \rho))$.
The second moment of this last passage time is proportional to
$g(0)$ with $g$ the scaling function in (\ref{eq:scalingform}). To
have the full function $g$ we have to determine the last passage time 
with an end point of order $n^{2/3}$ away from $(n/
\theta(\rho),n)$.

\setcounter{section}{4}
\section{The Baik and Rains analysis}\label{sec.e}
\setcounter{equation}{0}

Recently Baik and Rains \cite{BR} studied last passage percolation
with the above described boundary conditions. Unfortunately the case of an
exponential distribution is not yet accessible to their
techniques. Only the result of Johansson \cite{Joh} is
available which in our notation corresponds to $\rho_- =1$,
$\rho_+ = 0$. For this case he proves that
\begin{equation}
\lim_{t\to\infty}t^{-1/3}2^{4/3}(N_t-{\textstyle\frac14}t)
= - \xi_{\mbox{\scriptsize GUE}}
\end{equation}
in distribution with $\IP(\xi_{\mbox{\scriptsize GUE}} \leq x)=
  F_{\mbox{\scriptsize GUE}}(x)$.
$F_{\mbox{\scriptsize GUE}}$ is expressed in terms of a solution to the
  Painlev\'e II
equation. We will consistently use the notation of \cite{BR} for the
various distribution functions and will not repeat their
definitions here.

Baik and Rains study the geometric and Poisson last passage
percolation.
\medskip \\
(i)
{\it geometric}. The exponentially distributed random variables 
$w(i,j)$ are replaced by  $w_{q}(i,j)$, $0<q<1$.  $w_{q}(i,j)$, $i,j \geq 1$, 
has geometric
distribution with parameter $q$, $w_{q} (i,0)$, $i \geq 1$, has geometric
distribution with parameter $\alpha_+ \sqrt q$, and $w_{q}(0,j)$, $j
\geq 1$, has geometric distribution with parameter $\alpha_- \sqrt
q$. The random variables $\{ w_{q}(i,j)|\,i,j\geq 0\}$ are independent. Baik and Rains
study the distribution of the passage time $G_{q}(n,n)$ for large $n$
in dependence on $q$, $\alpha_+$, $\alpha_-$, where they allow
 $\alpha_+, \alpha_-$ being close to 1. Specifically they set
\begin{equation}
\label{eq:boundary}
\alpha_\pm=1-n^{-1/3}w_\pm
\end{equation}
and establish how the asymptotic distribution of $G_{q}(n,n)$ depends
on $w_{\pm}$. The particular case $\alpha_- = 0 = \alpha_+$ is
proved by Johansson \cite{Joh}, who can allow for an arbitrary
end point. In \cite{Barai} last passage percolation is investigated
for $\alpha_- = 0 = \alpha_+$
and the 
$w_{q}(i,j)$'s satisfying various reflection 
symmetries including the cases of interest here, 
reflection symmetric relative to the diagonal resp. relative to the
anti-diagonal. The
large $n$ distribution of $G_{q}(n,n)$ 
is established.

Following the scheme already explained we see that the geometric
case can be interpreted as a discrete time TASEP with waiting times
$w_{q}(i,j)+1$. In a single time
step every particle which has an empty site ahead can jump to the
right. They do so independently with probability $1-q$. Thus the
initial step, left density $\rho_-$ and right density $\rho_+$, is 
covered by
\cite{BR} and the periodic initial configuration,
$\ldots0101\ldots$, by \cite{Barai}. To handle the two-point
function, a small change in the boundary density as in
(\ref{eq:boundary}) should be translated to a small change in the
angle of the end point for the directed polymer away from the
diagonal. This seems to require a nontrivial generalization of the
known results. 

In the limit $q \rightarrow 1$ the random variables $(1-q)w_{q}(i,j)$ 
converge to $w(i,j)$ in distribution. Therefore the continuous time 
TASEP, time $t$,
is recovered from the discrete time TASEP at integer times $[t/ 
(1-q)]$ as $q \rightarrow 1$. \medskip\\
(ii) {\it Poisson}.  We consider the geometric last passage
percolation and interpret $w_{q}(i,j)$  as the number of points in the
square with corners $\sqrt q (i \pm \frac{1}{2}, j \pm
\frac{1}{2})$. Then in the limit $q\to 0$ such that $(u,v) =
\sqrt q\,(m,n)$ is fixed we obtain a Poisson point process which has
unit density in the rectangle $[0,u] \times [0,v]$, line density
$\alpha_+$ on $[0,u] \times \{0\}$, and line density $\alpha_-$ on
$\{0 \} \times [0,v]$. An allowed path, $\omega$, connects 
continuously $(0,0)$ to $(u,v)$, and is piecewise linear. 
Each linear segment has a slope $\theta$ such that $0 \leq \theta \leq
\infty$ and connects two Poisson points, except for the first 
and last piece.   $T(\omega)$
is the number of Poisson point which are transversed by $\omega$ and,
as before,
\begin{equation}
G_{\mathrm{Poisson}}(u,v)=\max_\omega T(\omega)\,.
\end{equation}
For $\alpha_- = 0 = \alpha_+$ the asymptotic distribution of
$G_{\mathrm{Poisson}}(v,v)$ is established in \cite{Ba dei jo}. Baik and Rains \cite{BR}
extend to arbitrary $\alpha_-, \alpha_+$ and allow for small
deviations of the boundary densities as in (\ref{eq:boundary}) with
$n^{-1/3}$ replaced by $v^{-1/3}$. Very recently
they investigate the case of reflection symmetry relative to the
diagonal, in particular $\alpha_- = \alpha_+$, and an extra line density
along the diagonal \cite{Barai2}.\vspace{-2mm}\\

The particle model behind the Poisson last passage percolation is
the PNG model \cite{Kru Spo, Prae Spo, Prae Spo2}. It consists of
point particles with velocity $\pm 1$. They annihilate each other
at a collision and are created in $\pm$ pairs with rate $1$. If
$\rho_+ (x,t)$ is the density of right movers, and $\rho_- (x,t)$ of left
movers, then the field $\phi (x,t) = \rho_+ (x,t) - \rho_- (x,t)$
is locally conserved and corresponds to $\eta_{t,j}$ of the TASEP.
More precisely the corresponding height function 
\begin{equation}
h(x,t) = N_{t} - \int_{0}^{x}dy\phi(y,t), 
\end{equation}
where as before $N_{t}$ is the time integrated current at the origin,
$N_{t}= \int_{0}^{t}ds(\rho_+ (0,s) + \rho_- (0,s))$. The height is 
related to the last passage time through $G(u,v)  = h(x,t)$ in 
distribution with $u= (t+x)/\sqrt{2}$, $v= (t-x)/\sqrt{2}$. 

In contrast to the geometric and exponential last passage percolation,
a change in the boundary densities $\alpha_{\pm}$ can be compensated 
by a variation of the end point. For example, for the square 
$[0,v]^{2}$ with boundary densities $\alpha_{\pm} = (1 - 
v^{-1/3}w)^{\pm 1}$, we
can stretch the $1$-axis by $1/\alpha_+$ and the $2$-axis by
$1/\alpha_-$. Then the boundary densities are $1$ and the end
point of the polymer is tilted by a distance $v^{2/3}$ away from
$(v,v)$, which is the quantity needed for the scaling form of the
$2$-point function. Thus for the PNG model the scaling form of
$\IE \big(\phi(x,t)\phi(0,0)\big)$ follows from the analogue Proposition
\ref{prop:two} for PNG and the asymptotic analysis in \cite{BR},
compare with Section \ref{7.2}.

The defect line along the diagonal is most easily visualized in
the growth version of PNG, where unit up-steps are particles with
velocity $-1$ and unit down-steps with velocity 1. If $\alpha_- =
0 = \alpha_+$, nucleation of pairs of steps are allowed only above
the ground layer $[-t, t]$. Extra Poisson points along the
diagonal with line density $\alpha$ correspond to a source at the
origin which nucleates at rate $\alpha$. The full PNG droplet
has no reflection symmetry at the diagonal. With reflection
symmetry one restricts oneself to the half droplet, where
nucleations only above the ground layer $[0,t]$ are allowed. In the
depinned phase the extra mass is incorporated in the droplet
without visible modification of the macroscopic shape. In the
pinned phase an extra pile of linear slope is created on top of the
PNG droplet.

\setcounter{section}{5}
\section{Scaling theory}\label{sec.f}
\setcounter{equation}{0}

In limit theorems one has to separate universal from
model-dependent features. To give the standard example: let
$\xi_j,\,j \in \IZ $, be a stationary mean zero sequence of
random variables which satisfies the central limit theorem as
\begin{equation}\label{CLT}
\lim_{n\to\infty}\IP\Big(\frac1{\sqrt{n}}\sum_{j=1}^n\xi_j\leq
  \sigma x\Big)=F_{\text{G}}(x)\,.
\end{equation}
Here $F_{\mathrm{G}}$ is the distribution function of a standard
Gaussian random variable. $F_{\mathrm{G}}$ is universal (within
well-studied limits), whereas the variance $\sigma^{2}$ depends on the
particular probability law for the $\xi_j$'s. In our example
$\sigma^{2}$ is easily computed as
\begin{equation}
\sigma^{2}=\sum_{j=-\infty}^\infty\IE\big(\xi_j\xi_0\big)\,.
\end{equation}

For the ASEP and similar growth models a corresponding scaling
theory is available which determines the model-dependent
longitudinal and transverse scales \cite{Kru Mea Ha}. To apply the theory one
has to know the stationary measures as labeled by their mean
density, $\rho$. The two basic quantities are
\begin{eqnarray*}
j(\rho)&=&\mbox{average current at density $\rho$}\,,
\nonumber\\
A(\rho)&=&\mbox{size of the density fluctuations in the stationary
  measure.}\nonumber
\end{eqnarray*}
If the density is regarded as the slope of a height function as in 
(\ref{eq:height}), then $A(\rho)$ is the roughness amplitude for $j \mapsto h(j)$
in the stationary measure, i.e. 
$A(\rho)=\lim_{j\to\infty}j^{-1}\IE\big(\big[h(j) - h(0) - \IE\big(h(j) 
-h(0)\big)\big]^{2}\big)$.

For the TASEP $j(\rho) = \mu_{\rho}(\eta_0(1-\eta_1)) =
\rho(1-\rho)$ and the height diffusion constant
\begin{equation}
A(\rho)=\lim_{j\to\infty}\frac1j\mu_\rho\big((h(j)-(1-2\rho)j)^2\big)
=4\sum_j(\mu_\rho(\eta_j\eta_0)-\rho^2)=4\rho(1-\rho)\,.
\end{equation}
For the discrete time TASEP, jump probability $1 - q$, the stationary
measure at density $\rho$ is a Markov chain with transition
probability $Q$, a $2\times2$--matrix which we write in the form
$Q_{00}=1-r$, $Q_{01}=r$,
$Q_{10}=s$, $Q_{11}=1-s$. $r,s$ are determined through  $(1-r)(1-s)= qrs$ and 
$(r+s)\rho=s$.
Then the average current
\begin{equation}
j(\rho)=\rho\,s
\end{equation}
and the roughness amplitude
\begin{equation}
A(\rho)=4\sum_{j=-\infty}^\infty\big(\IE_\rho\big(\eta_j\eta_0\big)-\rho^2 \big)
  =4\big(\rho(1-\rho)
    +2\sum_{j=1}^\infty(\rho(Q^j)_{11}-\rho^2)\big)\,.
\end{equation}
For the PNG model in the stationary measure the + , resp. $-$,
particles are Poisson distributed with density $\rho_+$, resp. 
$\rho_-$, 
satisfying the stationarity constraint $2\rho_+ \rho_- =1$. The conserved
density is $\rho = \rho_+ - \rho_-$ and the current
\begin{equation}
j(\rho)=\rho_++\rho_-=(2+\rho^2)^{1/2}.
\end{equation}
$ $From the Poisson property we conclude
\begin{equation}
A(\rho)=\rho_++\rho_-=(2+\rho^2)^{1/2}.
\end{equation}

$j'(\rho) = v(\rho)$ is the velocity of the density fluctuations
and $\lambda = j'' (\rho)$ is the KPZ coupling constant \cite{Kru Mea Ha}. 
In terms
of these quantities, the scale in the $h$-direction
is
\begin{equation}
\label{eq:j_ASEP}
\text{sign}\lambda({\textstyle \frac12}|\lambda|A^2)^{1/3}t^{1/3}
\end{equation}
and the scale in the transverse $j$-direction is
\begin{equation}
\label{eq:A_ASEP}
2^{4/3}(\lambda^2A)^{1/3}t^{2/3}\,,
\end{equation}
For the TASEP, since  $\lambda = 
-1$, the $h$-direction then comes
in units of $-2^{-1/3} (4\rho(1-\rho))^{2/3} t^{1/3}$, whereas the
$j$-direction comes in units of $4(\rho(1-\rho))^{1/3}t^{2/3}$.

In (\ref{eq:j_ASEP}) and (\ref{eq:A_ASEP}) we have fixed two 
dimensionless scale factors. Their role is most easily explained in 
our entry example, where we could have adopted a definition of the 
error function $F_{\mathrm{G}}$ such that the Gaussian has mean  
$1/2$. Then in (\ref{CLT}) $ \sigma x$ is to be replaced by $
\sqrt 2\sigma x $ independently of the value of $\sigma$. The 
particular convention is determined through the comparison with one 
single test 
case. In our context we follow \cite{BR} in the definition of the 
distribution functions, who rely on the 
established conventions for the 
Painlev\'{e} II Riemann-Hilbert problem. The test case ist the PNG 
model, which fixes the prefactors in (\ref{eq:j_ASEP}) to $2^{-1/3}$
and in (\ref{eq:A_ASEP}) to $2^{4/3}$.

So far proofs are available only for a short list of models. In all cases
the model-dependent parameters are such as to agree with the
scaling theory. In particular  the discrete time TASEP provides a
two-parameter ($\rho$ and $q$) test of the theory.

\setcounter{section}{6}
\section{Current fluctuations}\label{sec.g}
\setcounter{equation}{0}

There is little doubt that the results of Baik and Rains also hold
in the limit of an exponential distribution when $q\to1$. Strictly
speaking our results are conjectures, except for $\rho_- = 1$,
$\rho_+ =0$ \cite{Joh}. We discuss the list of our examples. Some
of them have been announced in \cite {Prae Spo2}, where also
numerical plots of $F_{\text{GUE}}$, $F_{\text{GOE}}$, and $F_0$ are given.

\subsection{Initial step density}

\begin{figure}[t]
  \begin{center}
    \mbox{\epsfig{file=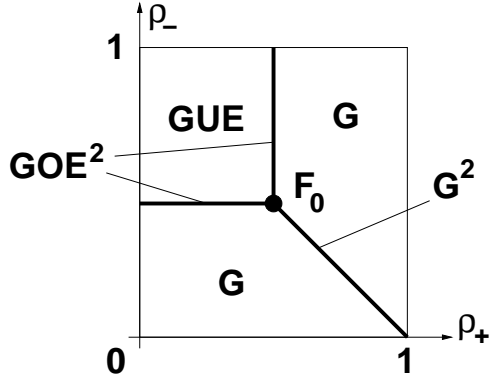,height=5cm,angle=0}}
  \end{center}
  \caption{\label{fig:phase}\em Fluctuation phase diagram for the TASEP.}
\end{figure}

Fluctuation results for $N_t$ are most easily summarized in the
diagram of Figure \ref{fig:phase}.
We first remind the reader of the limit measure $\mu_{\infty}$ for
$\{\eta_{t,j}$, $|j| \leq N \}$, $N$ arbitrary, $t\to\infty$
\cite{Ligg}. The upper left corner is the maximal current phase
with $\mu_{\infty} = \mu_{1/2}$. In the lower left $\mu_{\infty}=
\mu_{\rho_-}$ and in the upper right $\mu_{\infty} =
\mu_{\rho_+}$. Along the line $\rho_- + \rho_+ = 1$ the long time
limit is the superposition $\mu_{\infty} = \frac{1}{2}
\mu_{\rho_-} + \frac{1}{2} \mu_{\rho_+}$. In particular, we have
$\IE\big(N_t\big)/t\to \mu_{\infty}(\eta_0 (1- \eta_1))$ as
$t\to \infty$.

To understand the fluctuations of $N_t$ we use the directed
polymer picture and recall that the polymer starts at $(0,0)$ and ends 
at $(n,n)$. If $\rho_- < 1 - \rho_+$, $\rho_- < \frac{1}{2}$,
then the polymer stays a finite fraction of steps at the right
edge. Since for the edge steps the passage time is a sum of
independent exponentials, its fluctuations are of order $\sqrt t$
and Gaussian (G). At some point the polymer must enter the
bulk and we expect subleading corrections of order $t^{1/3}$, 
\begin{equation}N_t\cong\rho(1-\rho)t+t^{1/2}\xi_{\mbox{\scriptsize
      G}}-t^{1/3}\xi_{\mbox{\scriptsize GUE}}
\end{equation}
up to prefactors, with either $\rho = \rho_-$ or $\rho = \rho_+$. If $\rho_- +
\rho_+ = 1, \rho_- < \frac{1}{2}$, then the polymer has  a choice 
between the 
left and right edge and the fluctuations of $N_t$ are the maximum
of two independent Gaussians (G$^2$). Recall that if $\xi_i$ has
distribution function $F_i$, $i = 1,2$, then their maximum has the
distribution function $F_1 F_2$ provided $\xi_1, \xi_2$ are
independent, hence our notation.

If $\rho_- > \frac{1}{2}$, $\rho_+ < \frac{1}{2}$, then it does
not pay for the polymer to stay at neither edge. The fluctuations
must be as for the Johansson case of trivial edges, $\rho_- = 1$,
$\rho_+ = 0$, which have $F_{\mbox{\scriptsize GUE}}$ as
distribution function. Thus the difficult cases are the two
critical lines $\rho_- = \frac{1}{2}$, $\rho_+ < \frac{1}{2}$, resp.
$\rho_- > \frac{1}{2}$, $\rho_+ = \frac{1}{2}$, and the critical
point $\rho_- = \frac{1}{2} = \rho_+$, which corresponds to the
Bernoulli $\frac{1}{2}$ initial measure. On the line $\rho_- =
\frac{1}{2}$ $(\rho_+ = \frac{1}{2})$ the directed polymer stays
order $n^{1/3}$ at the left (right) edge, whereas at the critical
point it makes a choice between the two edges. According to
\cite{BR} the fluctuations of $N_t$ are of order $t^{1/3}$ and are
$\mbox{GOE}^2$ distributed on the critical lines, $F_0$
distributed at the critical point.

From the point of view of the last passage percolation the
end point $(n,n)$ is somewhat special and one might consider more
generally the end point $(m,n)$ with the slope $\theta = n/m$
fixed. To have a firm link to the TASEP we will use the
parameterization (\ref{eq:height}), in which case the quantity of
interest is $h_t([yt])$, $|y| < 1$, $[\cdot]$ denoting the integer part. 
$h_t(0) = 2 N_t$ is the
particular case just explained. The hydrodynamic theory \cite{Re,
Sep1} establishes the law of large numbers for $h_t([yt])$ with the
result
\begin{equation}
\lim_{t\to\infty}\frac1t h_t([yt])=\bar h(y)
\end{equation}
almost surely for $|y| < 1$. The limit $\bar{h}$ depends on
$\rho_-, \rho_+$ and is given as follows: If $\rho_- < \rho_+$,
then
\begin{eqnarray}
\label{eq:meanheight1}
\bar h(y)=\left\{
  \begin{array}{lcr}
    (1-2\rho_-)y+2\rho_-(1-\rho_-)&\mbox{for}&y\leq y_c \,, \\
    (1-2\rho_+)y+2\rho_+(1-\rho_+)&\mbox{for}&y> y_c\\
  \end{array}
\right.
\end{eqnarray}
with $y_c = \big(\rho_+ (1- \rho_+) - \rho_-(1- \rho_-)\big) / (\rho_+ -
\rho_-)$. If $\rho_- > \rho_+$, then
\begin{eqnarray}
\label{eq:meanheight2}
\bar h(y)=\left\{
  \begin{array}{lcc}
    (1-2\rho_-)y+2\rho_-(1-\rho_-)&\mbox{for}&y\leq 1-2\rho_- \,,\\
    \frac12(y^2+1)&\mbox{for}&1-2\rho_-<y\leq1-2\rho_+ \,, \\
    (1-2\rho_+)y+2\rho_+(1-\rho_+)&\mbox{for}&1-2\rho_+<y\,.\\
  \end{array}
\right.
\end{eqnarray}

We are interested in the fluctuations of
\begin{equation}
h_t([yt])-t\bar h(y)
\end{equation}
for fixed $y$ in the limit $t\to\infty$. If $\bar{h}$ is linear around
$y$, then the fluctuations are order $\sqrt t$ and Gaussian. If
$\bar{h}$ has a cusp at $y$ (and is linear on both sides), then
the fluctuations are order $\sqrt t$ and $\mbox{G}^2$. On the
other hand, if $\bar{h}$ has nonzero curvature at $y$ then the
fluctuations are order $t^{1/3}$ and $\mbox{GUE}$.
The critical lines, $\text{GOE}^2$, correspond to a $y$
where at $\bar{h}(y)$ the curved piece joins the linear piece.
Finally the critical point, $F_{0}$, is the merger of the two critical
lines.

We summarize our findings in the form of a (well-founded)
conjecture. $\IP_{\rho_-, \rho_+}$ refers to the TASEP with
$\mu_{\rho_-, \rho_+}$ as starting measure. $h_t (j)$ is defined
in (\ref{eq:height}), $\bar{h}$ in (\ref{eq:meanheight1}),
(\ref{eq:meanheight2}).
\begin{conjecture}\label{conj}
{\em(G)} Let either $\rho_- < \rho_+$ and $y > y_c$ or $\rho_- >
\rho_+$ and $y > 1-2 \rho_+$. Then
\begin{equation}\label{conjec1}
\lim_{t\to\infty}\IP_{\rho_-,\rho_+}\Big(t\bar
  h(y) 
 - h_t([yt])\leq \big(4\rho_+(1-\rho_+)(y-1+2\rho_+)t
 \big)^{1/2}x\Big)
=F_{\mbox{\em\scriptsize G}} (x)\,.
\end{equation}
Let either $\rho_- < \rho_+$ and $y < y_c$ or $\rho_- > \rho_+$
and $y < 1-2 \rho_-$. Then
\begin{equation}\label{conjec2}
\lim_{t\to\infty}\IP_{\rho_-,\rho_+}\Big(t\bar
  h(y) - h_t([yt])\leq \big(
  4\rho_-(1-\rho_-)(-y+1-2\rho_-)t\big)^{1/2}x\Big)
=F_{\mbox{\em\scriptsize G}} (x)\,.
\end{equation}
{\em(G$^2$)} Let $\rho_- < \rho_+$ and $y = y_c$, then
\begin{eqnarray}\label{conjec3}
\lim_{t\to\infty}\IP_{\rho_-,\rho_+}\Big(t\bar
  h(y) - h_t([yt])\leq ((\rho_+ - \rho_-) t)^{1/2}x\Big)&&\nonumber\\
&&\hspace{-7cm}=F_{\mbox{\em\scriptsize G}}
\big((4\rho_+(1-\rho_+))^{-1/2}x\big)
F_{\mbox{\em\scriptsize G}}
\big((4\rho_-(1-\rho_-))^{-1/2}x\big)\,.
\end{eqnarray}
{\em(GUE)} Let $\rho_- > \rho_+$ and $1-2 \rho_- < y < 1-2 \rho_+$.
Then
\begin{equation}\label{conjec4}
\lim_{t\to\infty}\IP_{\rho_-,\rho_+}\Big(t\bar
  h(y) - h_t([yt])\leq 2^{-1/3}(1-y^2)^{2/3}t^{1/3}x\Big)
=F_{\mbox{\em\scriptsize GUE}} (x)\,.
\end{equation}
{\em(GOE$^{2}$)} Let $\rho_- > \rho_+$ and either $y = 1-2 \rho_-$
or $y = 1-2 \rho_+$. Then 
\begin{equation}\label{conjec5}
\lim_{t\to\infty}\IP_{\rho_-,\rho_+}\Big(t\bar
  h(y) - h_t([yt])\leq 2^{-1/3}(1-y^2)^{2/3}t^{1/3}x\Big)
=F_{\mbox{\em\scriptsize GOE}} (x)^2\,.
\end{equation}
{\em(F$_{0}$)} Let  $\rho_- = \rho = \rho_+$ and $y = 1-2 \rho$. Then
\begin{equation}\label{conjec6}
\lim_{t\to\infty}\IP_{\rho_-,\rho_+}\Big(t\bar
  h(y) - h_t([yt])\leq 2^{-1/3} (1-y^2)^{2/3}t^{1/3}x\Big)
=F_{0} (x)\,.
\end{equation}

$F_{\text{\em G}}$ is the distribution function of a standard
normal distributed random variable. $F_{0}$ and the Tracy-Widom distribution
functions $F_{\text{\em GUE}}$, $F_{\text{\em GOE}}$ are defined
in \cite{BR}.
\end{conjecture}
The Gaussian case (G) with $\rho_- =  \rho_+$ is 
proved in \cite{FF} and the case (GUE)
with $\rho_- = 1$, $\rho_+ = 0$ in \cite{Joh}. 

In (\ref{conjec1}) to (\ref{conjec6}) the random variable 
$-(h_t([yt]) - t\bar h(y)) $ appears because of the inversion from the
passage time $G(m,n)$ to the height. In each case we have subtracted the
asymptotic mean as computed from the hydrodynamic theory. For the
variances of the Gaussians we have to determine how many
steps, on the scale $n$, the directed polymer stays at either
edge. For the prefactor $2^{-1/3}(1-y^2)^{2/3}$ in (\ref{conjec4}) to
(\ref{conjec6}) we either take the limit of the discrete time
TASEP or use the scaling theory of Section \ref{sec.f}. Because of 
the inversion, also the lower and upper tails are interchanged. For 
example $F_{\mathrm{GUE}}(x)$ has the lower tail
$\exp(- |x|^{3}/12)$
and the upper tail $\exp(- 4|x|^{3/2}/3)$.

\subsection{Stationary two-point function}\label{7.2}

The starting measure is Bernoulli $\rho$ and $\IP_{\rho}$ is the
corresponding path measure of the stationary TASEP, $0 < \rho <
1$. Density fluctuations propagate with velocity $1 - 2\rho$. Therefore 
the anomalous fluctuations appear in $h_t([(1 - 2\rho)t])$, compare 
with (\ref{conjec6}). For the full two-point function we need the 
height distribution a distance of order $t^{2/3}$ away, i.e. at $(1 - 2\rho)t
+ 4(\rho(1-\rho))^{1/3}t^{2/3}w$ for arbitrary $w$, compare with (\ref{eq:A_ASEP}).
\begin{conjecture}
We have
\begin{eqnarray}
\label{eq:monster}
&&\lim_{t\to\infty}\IP_\rho\Big(- h_t([(1-2\rho)t
    +4(\rho(1-\rho))^{1/3}t^{2/3}w])
    +2\rho(1-\rho)t
  \nonumber\\
&& \hspace{1.6cm}+(1-2\rho)\big((1-2\rho)t +4(\rho(1-\rho))^{1/3}t^{2/3}w\big)
\leq 2(\rho(1-\rho))^{2/3}t^{1/3}x\Big)
\nonumber\\
&&\hspace{1cm}= F_w(x)\,.
\end{eqnarray}
\begin{figure}[t]
  \begin{center}
    \mbox{\epsfig{file=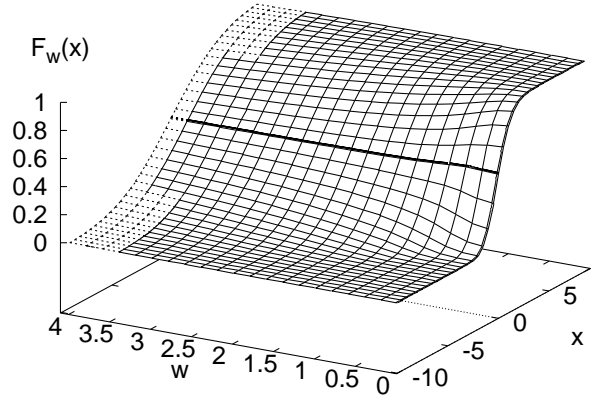,height=7cm,angle=0}}
  \end{center}
\vspace{-1cm}
  \caption{\label{fig:Fw}\em
    A 3d-plot of the distribution functions $F_w$
    for $w>0$.}
\end{figure}
The distribution function $F_w(x)$ is defined in \cite{BR} and denoted
there by $H(4w^{2} + x; w, -w)$. The distribution function $F_{0}$ appears 
also in (\ref{conjec6}). 
\end{conjecture}
Note that $\int d F_w(x)x = 0$, since we have subtracted the
average of $h_t$. The prefactors of $x$ and  $w$ 
are infered from the scaling theory of Section \ref{sec.f}. 
For the PNG model, the analogue of 
(\ref{eq:monster})
follows from  \cite{BR}. If the convergence
(\ref{eq:monster}) holds
also for the second moments, then
\begin{eqnarray}
\label{eq:yamonster}
&&\IE_\rho\big(\big[h_t([(1-2\rho))t
+4(\rho(1-\rho))^{1/3}t^{2/3}w])
    -2\rho(1-\rho)t
\nonumber\\
&& \hspace{2.6cm} -(1-2\rho)((1-2\rho)t
+4(\rho(1-\rho))^{1/3}t^{2/3}w)\big]^2\big)
\nonumber\\
&& \hspace{0.4cm}\cong 4(\rho(1-\rho))^{4/3}t^{2/3}\int
dF_w(x)x^2
\end{eqnarray}
for large $t$. Setting
\begin{equation}
g(w)=\int dF_w(x)x^2
\end{equation}
and taking second derivatives in (\ref{eq:yamonster}), we conclude the
scaling form of (\ref{eq:scalingform}).

\begin{figure}[h]
  \begin{center}
    \mbox{\epsfig{file=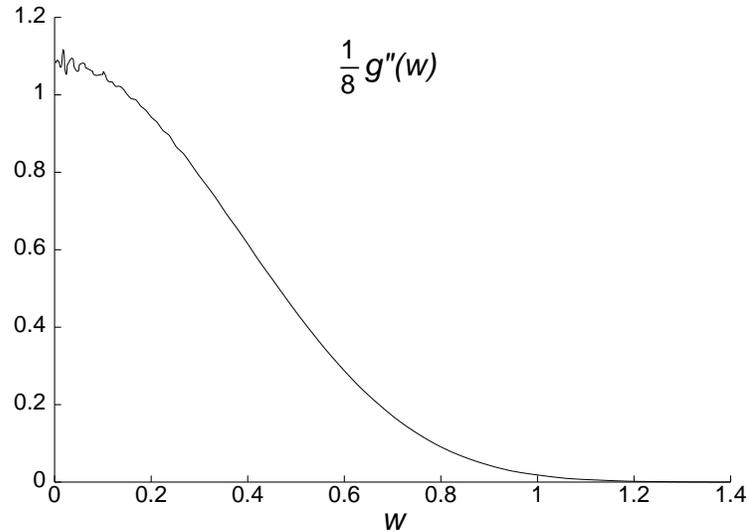,width=7cm,angle=270}}
  \end{center}
  \caption{\label{fig:gpp}\em The second derivative of the 
  second moment of $F_w$ vs. $w$.}
\end{figure}

$F_w(x)$ is the solution of a first order
partial differential equation with suitable boundary conditions. A
numerical plot is provided in Figure \ref{fig:Fw}.
In essence it shows how the distribution broadens for $w
\rightarrow \infty$, reflecting the cross over to the Gaussian
fluctuations as one moves away from the characteristic $\{j = (1-2
\rho)t\}.$ The second moment, $g(w)$, is symmetric and increases linearly 
as $4|w|$ for large $|w|$.
In Figure 3 we plot the scaling function $g''$ as determined from 
the numerical evaluation of the second moment of $F_{w}$ at various 
values of $w$. The oscillations at small $w$ result from numerical 
imprecisions for $F_{w}$.   In \cite{Bei Ku
Spo} an approximate nonlinear integral equation for $g''$ has been 
derived, which was then solved numerically in \cite{Hwa Frey}.
The approximate $g(0)$ differs from the exact one by an order of
$5\%$. From the available data a more accurate comparison does not
seem to be feasible.

\subsection{Semi-infinite system}

We restrict to the half-lattice $\IN$ and insert with rate $\alpha$
particles at site $1$, respecting the constraint $\eta_1=0,1$. The
initial measure is Bernoulli $\mu_\rho$. The last passage percolation
representation lives in the lower triangle $\{(i,j)|\, 0 \leq j\leq i\}$. Along the
diagonal the $w(j,j)$ are exponential with rate $\alpha$ and at the
lower edge the $w(j,0)$ are exponential with rate $1-\rho$. By
maximizing the passage time on scale $n$ we obtain the same phase
diagram as for the critical step. However the fluctuations in $N_t$,
now the number of particles injected up to time $t$, are modified.
In Figure 4 we summarize the findings \cite{Barai2}  which could be written 
more formally as in Conjecture \ref{conj} and so far have been proven
only for the  
Poisson case, i.e. the semi-infinite PNG model with a source at the
origin.
\begin{figure}[b]
  \begin{center}
    \mbox{\epsfig{file=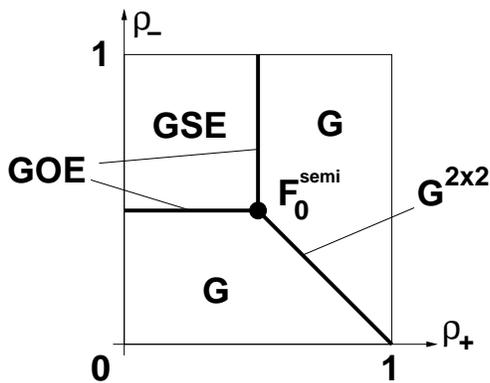,height=5cm,angle=0}}
  \end{center}
  \caption{\label{fig:phasesemi}\em Fluctuation phase diagram for the
    semi-infinite TASEP.}
\end{figure}
GSE is the distribution of the largest eigenvalue of a symplectic
Gaussian random matrix. $F_0^{\text{semi}}$ is a novel distribution
 and given by
\begin{equation}
  \label{eq:novel}
  F_0^{\text{semi}}(x)=\big(1+{\textstyle\frac12}(-v(x)+u(x))
  (x+2u'(x)+2u(x)^2)\big)F(x)E(x)^3
\end{equation}
in the notation of \cite{BR}.
On the anti-diagonal a new feature appears. Recall that for the
infinite system the distribution G$^2$ arose from the choice between
the left and right edge. Here the directed polymer may stay for an
arbitrary length at the lower edge and then switch to the
diagonal. Optimizing over this one parameter family of paths yields
for the passage time a distribution which is given by the largest
eigen value of a $2\times2$ GUE matrix, denoted by G$^{2\times2}$ in
Figure \ref{fig:phasesemi}.
 
\subsection{Defect lines, depinning}

For a change let us consider the Poisson last passage percolation. As 
already explained, in the square $[0,v]^{2}$ we have Poisson points with 
density 1. The directed polymer starts at $(0,0)$ and ends at 
$(v,v)$. We add extra Poisson points along the diagonal with line 
density $\alpha$. Of primary interest is the last passage time, 
$\lim_{{v \to \infty}}v^{-1}G_{\alpha}(v) = \tau(\alpha)$. We know 
that $\tau(0) = 2$. The critical $\alpha_{c}$ is defined as 
the largest $\alpha$ such that $\tau(\alpha) = 2$. If $\alpha > 
\alpha_{c}$, the polymer is pinned and typically returns after a 
length $\ell_{\|}(\alpha)$ to the diagonal. Between two returns 
the polymer makes an excursion of size $\ell_{\bot}(\alpha)$ away 
from the diagonal. The depinning transition is governed by 
$\tau(\alpha) , \ell_{\|}(\alpha),\ell_{\bot}(\alpha)$ as $\alpha 
\downarrow \alpha_{c}$. If  $\ell_{\bot}(\alpha_{c}) < \infty$, the 
transition is first order, while  $\ell_{\bot}(\alpha) \to \infty$ as
$\alpha 
\downarrow \alpha_{c}$ signals as second order transition.

Since there is considerable interest from the physics of disordered 
systems, a phenomenological theory has been developed, which after 
some controversy seems to be widely accepted \cite{HN,Tang}. We explain the 
predictions and first recall the pure problem with the path measure
\begin{equation}\label{eq:BB}
    (Z_{t})^{-1} \IP^{BB} \exp\big[ -\beta \int _{0}^{t} ds 
    V(x_{s})\big] \,.
\end{equation}
Here $x_{s} \in \IR^{d}$ is a Brownian path which has the a priori 
weight given by the Brownian bridge $\IP^{BB}$ with start and end 
point $0$. $V$ is a pinning potential, which by a suitable choice of 
units is simply $V(x) = 0$ for $|x| \geq 1$ and $V(x) = - 1$ for $|x| < 
1$. $Z_{t}$ is the normalizing partition function. The analogue of 
$\tau(\alpha)$ is the ground state energy $e(\beta) = \lim_{t \to 
\infty} - t^{-1}\log Z_{t}$, $e(0) = 0$. For $d=3$ a certain minimal 
strength is required to pin the Brownian motion. Thus $\beta_{c} > 
0$ and $e(\beta) \simeq - (\beta - \beta_{c})^{2} $ for $
\beta \geq \beta_{c}$. However for $d=2$, Brownian motion is 
null-recurrent and $\beta_{c} = 0$. Just a tiny bit of attraction 
suffices to pin the polymer. Correspondingly $e(\beta)$ has an 
essential singularity as $e(\beta) \simeq - e^{-1/\beta}$, 
$\ell_{\|}(\beta) \simeq   e^{1/\beta}$, $\ell_{\bot}(\beta) \simeq 
\ell_{\|}(\beta)^{1/2}$ for small $\beta$.

The claim is that a directed polymer in $1+1$ dimensions in a random 
potential is pinned in essence as a directed polymer in $2+1$ 
dimensions with zero bulk potential. Specifically, any attraction 
along the diagonal will pin the polymer, i.e. $\alpha_{c} = 0$,  
the passage time $\tau(\alpha) \simeq  2 + e^{-1/\alpha}$, and the 
longitidonal excursions 
$\ell_{\|}(\alpha) \simeq   e^{1/\alpha}$ with the usual link
$\ell_{\bot}(\alpha) \simeq 
\ell_{\|}(\alpha)^{2/3}$ for small $\alpha$. No rigorous result seems 
to be available.

Transcribed to the slow bond problem, the prediction is $r_{c} = 1$ 
with an essential singularity for $j_{\infty}(r)$ at $r = 1$.

To model the case where the Poisson points are reflection symmetric 
relative to the diagonal, we add in (\ref{eq:BB}) a hard wall and 
replace $V$ by $V_{\mathrm{hw}}$, $V_{\mathrm{hw}}(x_{1}, x_{\bot})=
V(x)$ for $x_{1}\geq 0$, $V_{\mathrm{hw}}(x) = \infty$ for $x_{1} < 0$.
The entropic repulsion shifts to $\beta_{c} > 0$ even for $d=1,2$. 
Therefore the prediction of the phenomenological theory is 
$\tau(\alpha) \simeq 2 + c(\alpha - \alpha_{c})^{2}$, 
$\ell_{\|}(\alpha) \simeq (\alpha - \alpha_{c})^{-2}$, and 
$\ell_{\bot}(\alpha) \simeq 
\ell_{\|}(\alpha)^{2/3}$ for $\alpha$ close to $\alpha_{c}$, $\alpha 
\geq \alpha_{c}$. This is in agreement with the exact result
$\alpha_{c}= 1/\sqrt{2}$ and $\tau(\alpha) = 2$ for $0 \leq \alpha 
\leq \alpha_{c}$, $\tau(\alpha) = \sqrt{2}(\alpha + (1/2\alpha))$ 
for $\alpha_{c} \geq \alpha$, which slightly above $\alpha_{c}$ yields
$\tau(\alpha) \simeq 2 + 2(\alpha - \alpha_{c})^{2}$
\cite{Barai}.

\setcounter{section}{8}
\section{Conclusions}\label{sec.h}
\setcounter{equation}{0}

It is rather surprising to have now, after more than 20 years of 
study, a technique available through which universal scaling 
functions can be computed, in some cases even very explicitly. The 
main lesson to be drawn is that the universal distribution functions 
on the scale $t^{1/3}$ depend on 
the type of initial conditions. It would be of interest to understand
whether our list is already complete.

Despite progress there are several obvious elements missing. Firstly 
for 
the transverse fluctuations of the polymer we do not have such a fine 
information as on the passage time (= energy of the polymer) \cite{joh}.
More importantly, only single distributions, like the passage time 
with given end points, can be handled. To have a more detailed 
understanding of the energy landscape joint distributions must be 
studied, like the joint distribution of the first passage times
$G(m_{1},n_{1}),G(m_{2},n_{2})$ refering to two distinct end points.

\end{document}